\documentclass[aps,floatfix,twocolumn,nofootinbib]{revtex4}
\usepackage{epsfig}
\begin{document}
\def\twoplus{$2^+$~}
\def\numA{519~}
\def\numexp{557~}
\def\numexpQ{328~}
\def\numQ{318~}
\def\numstaticq{98~}
\title{Systematics of the first \twoplus excitation with the Gogny interaction}
\author{ G.F. Bertsch$^{1}$, M. Girod$^{2}$, S. Hilaire$^{2}$, J.-P. Delaroche$^{2}$, H. Goutte$^{2}$ and S. P\'eru$^{2}$
}
\affiliation{
$^{1}$Department of Physics and Institute of Nuclear Theory,
Box 351560\\
University of Washington
Seattle, WA 98915 USA\\
$^{2}$ CEA/DAM Ile de France, DPTA/Service de Physique Nucl\'eaire, BP
12, 91680 Bruy\`eres-le-Chatel, France\\
}

\def\be{\begin{equation}}
\def\ee{\end{equation}}
\def\rme{$\langle 2 || \hat{Q}_2 || 0 \rangle$}
\def\static{$\langle 2 | Q | 2\rangle$}
\begin{abstract}

We report the first comprehensive calculations of \twoplus excitations 
with a microscopic theory applicable to over 90\% of the known nuclei.
The
theory is based on a quantal collective Hamiltonian in five
dimensions in which the potential energy and the tensor
of inertia are obtained from constrained triaxial
Hartree-Fock-Bogoliubov calculations. The only parameters
in theory are those of the finite-range, density-dependent Gogny
D1S interaction. The following properties of the lowest \twoplus
excitations are calculated: excitation energy, reduced
transition probability, and
spectroscopic quadrupole moment. \numA nuclei are included in our 
survey, comprising all but 38
of the known nuclei tabulated by Raman et al. \cite{ra02}. We find that the theory is very
reliable to classify the nuclei by shape. Quantitatively the
performance of the theory in deformed nuclei is excellent:
average excitation energies and transition quadrupole moments
are within 5\% of the experimental values, and the dispersion
about the averages are roughly 20\% and 10\%, respectively.  The
performance is not as good on spherical and soft nuclei. 
Including all nuclei in the performance evaluation, the average
transition quadrupole moment is 11\% too high and the dispersion
about the mean is $\sim20$\%.  For the energies, the average is 13\%
too high and the dispersion is $\sim40$\%.
\end{abstract}
\maketitle

\section{Introduction}

A framework for a comprehensive theory of nuclear structure has
been discussed for a long time \cite{ri80}, but up to now
there has been no systematic evaluation of the accuracy or reliability
of different methods used and their underlying energy functionals.
Methodologies based on self-consistent mean field theory or 
density functional theory have been extended in 
different ways with many of their details dependent on the energy 
functional's form. Except for one new study \cite{sa06},
the extensions to treat properties of excited states
have been tested only in limited regions of nuclei.
In this work we
examine the properties of the lowest \twoplus excited state over
the periodic table as a whole, using a methodology and energy
functional that has been
quite successful in previous studies of strongly deformed nuclei
\cite{de06,li99,go05} and soft nuclei~\cite{Bo88,Fl04}.  

The present theory uses the Generator Coordinate
Method with the Gaussian Overlap Approximation (GCM+GOA) to 
construct a collective Hamiltonian.  The elements of the theory
are well-known in nuclear physics ~\cite{ri80}.   One starts with the
Constrained Hartree-Fock-Bogoliubov (CHFB) theory of the
potential energy surface, and constructs a collective
Hamiltonian from the potential energy surface and the
information about the kinetic energy operator obtained from the
wave functions on that surface. The finite-range
density-dependent Gogny interaction D1S is used throughout; the
only parameters in the theory are those of D1S~\cite{de80,be91}.
The microscopic collective Hamiltonian (5DCH) is formally similar to 
the Bohr Hamiltonian. It has six kinetic terms, associated with three
rotational moments of inertia and three mass parameters stemming
from the fluctuations of axial and triaxial deformations. The
rotational moments of inertia are obtained using the
Thouless-Valatin prescription from the CHFB solutions in the
presence of a small rotational field.
We emphasize that there are no adjusted parameters in
the present treatment of the rotational moments of inertia.
The mass parameters are calculated by the
Inglis-Belyaev formula, which only requires local information
about the CHFB solutions on the grid points of a mesh in
deformation space. This is sufficient information to construct
the collective Hamiltonian, but not enough to compute matrix
elements.  For that, one needs the overlaps between states of
different deformations.  This is also computed using a local
approximation \cite{li99}.  The local approximations are quite
accurate for heavy open shell nuclei, but break down when
applied to doubly magic nuclei, because the overlap between CHFB
states is not sharp enough~\cite{ri80}. We therefore exclude
those nuclei from this study. The properties we discuss are the
excitation energy $E$, the transition quadrupole moment
\rme~between the ground and the excited states, and the
spectroscopic quadrupole moment $Q(2^+)$ of the excited state.

\section{Results}

There are \numexp even-even nuclei with known \twoplus
excitation energies as of compilation by Raman {\it et al.} in
2001 \cite{ra02}.  Their excitation energies span more than two
orders of magnitude, presenting a very substantial challenge to
any global theory of nuclear structure.  In our study here we
limit our scope somewhat by excluding the very light nuclei ($Z$
or $N < 8$), for which mean field theory is least justified. 
This eliminates 16 nuclei.  Also, the mapping of the CHFB to the
collective Hamiltonian becomes problematic for rigid spherical
nuclei such as the doubly magic ones.  An additional 23 nuclei
have been eliminated for that reason, leaving \numA nuclei in
the present study.  We first discuss the energies and then the
quadrupole properties of the nuclei

\subsection{Excitation energies}

\begin{figure}
\includegraphics [width = 6cm]{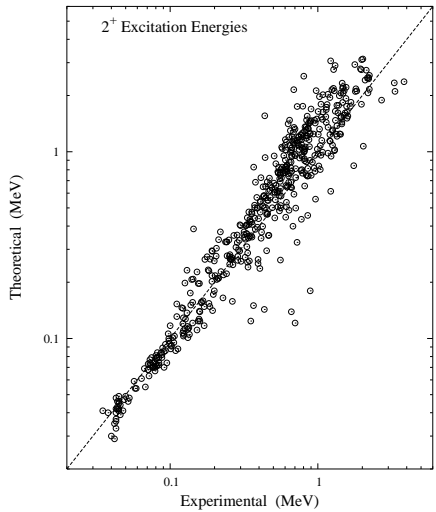}
\caption{Scatter plot of \numA even-even nuclei as a function of their
experimental and theoretical \twoplus excitation energies. }
\label{fi:e_scatter}
\end{figure}

Fig.~\ref{fi:e_scatter} shows a scatter plot comparing experimental
and theoretical excitation energies.  The points follow the
diagonal line fairly well with some scatter that varies in
extent over the different excitation energy regimes.  The lowest
energies are for the heavy, strongly deformed actinide nuclei;
the theoretical energies here are the most accurate (on a
logarithmic as well as absolute scale).  At excitation energies
of 1 MeV and higher, the theory has only a qualitative
predictive power, with errors ranging to a factor of 2 and
larger.  In the middle the theory improves but one can see a few
nuclei far from the diagonal.  They correspond to neutron-deficient
isotopes of Hg and Pb, where there is a near-degeneracy of 
weakly deformed oblate and well-deformed prolate structures. 

\begin{figure}
\includegraphics [width = 6cm]{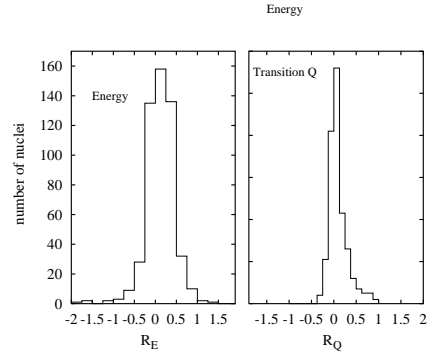}

\caption{Histogram of the logarithmic errors  $R_E,R_Q$ of the present
theory. Left panel: energies of 
the first excited \twoplus states for \numA of the \numexp nuclei whose
excitation energies are tabulated in ref. \cite{ra02}; right panel:
matrix elements $\langle 2 || \hat{Q}_2 || 0 \rangle$
for the transition between the ground and first excited state for
\numQ of the \numexpQ nuclei whose $B(E2)$ values
are tabulated in ref. \cite{ra02}.
}

\label{fi:e-hist}
\end{figure}

To make a quantitative measure of the theoretical accuracy, we
compare theory and experiment on a logarithmic scale,
examining the statistical properties of the quantity
$
R_E = \log (E_{\rm th}/E_{\rm exp})$.
Here $E_{\rm th}$ and $E_{\rm exp}$ are the theoretical and experimental
excitations energies, respectively.  A histogram of distribution of the
$R_E$'s is shown in Fig.~\ref{fi:e-hist}.  One can
see that there is a bias to positive values of $R_E$, i.e.
an overprediction of the excitation energy.  For the set
of \numA nuclei, the average is
$\bar R_E =0.12$.  Thus, there is a systematic bias to overestimate
the excitation energy by  12\%.  

The width of the distribution is the important quantity to
determine the accuracy and reliability of the theory.  One can
see from the histogram that the peak in the distribution extends
from $R_E$ about
-0.25 to +0.50, with a small tail going much farther from zero.  
A single number cannot be adequate to express the width of such 
a distribution, but for purposes of future comparisons with other
theories we report the root mean
square deviation of $R_E$ about its mean.  This measure comes out
to $\sigma_E \equiv \langle \Delta R_E^2\rangle^{1/2} = 0.33$. This implies
that typical errors are around -30\% on the low side to
+40\% on the high side after correcting for the systematic bias.
While these results may seem disappointing, one should remember that
present predictions are from a global theory with no parameter adjustment.

\subsection{Quadrupolar properties}

\begin{figure}
\includegraphics [width = 6cm]{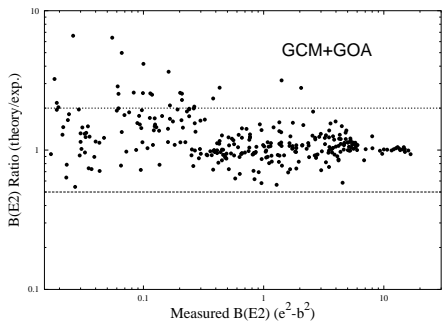}
\caption{Experiment compared to theory for the $B(E2, 0^+\rightarrow 2^+)$
for the nuclei tabulated in
ref. \cite{ra02}.  This graph may be directly compared with their Fig. C. 
Values within the lines are within a factor of two
of experiment.  Of the 306 cases shown here,  93\% are within the error
band.   This is superior to their ``GLOBAL" phenomenological fit and is 
much better than the theoretical models they consider.
\label{fi:q_scatter}
}
\end{figure}

The compilation of Raman {\it et al.} includes \numexpQ measured
quadrupole transition rates.  Of these, \numQ met the conditions
for applying the 5DCH theory.  The comparison between
theoretical and experimental reduced transition rates $B(E2, 0^+
\rightarrow 2^+)$ are shown as a scatter plot in
Fig.~\ref{fi:q_scatter}.
One sees that the theory is quite accurate for the largest values; these
are the actinide nuclei which are both heavy and strongly deformed.

For a quantitative measure of the accuracy of the theory, we define
$
R_Q = \log \left(\langle 2 || \hat{Q}_2 || 0 \rangle_{th}/ \langle 2 || \hat{Q}_2 || 0
\rangle_{exp}\right),
$
the logarithm of the ratio of the transition matrix elements. 
The distribution of $R_Q$ values is plotted as a histogram in
the right panel of 
Fig.~\ref{fi:e-hist}.  We see that the distribution is narrow
and centered close to zero.  The average of $R_Q$ values is
0.10, corresponding to a matrix elements 11\% too large on
average.  The dispersion $\sigma_Q$ is 0.21, corresponding to a
range of $+20/-19$ \% about the center value.

\begin{figure}
\includegraphics [width = 6cm]{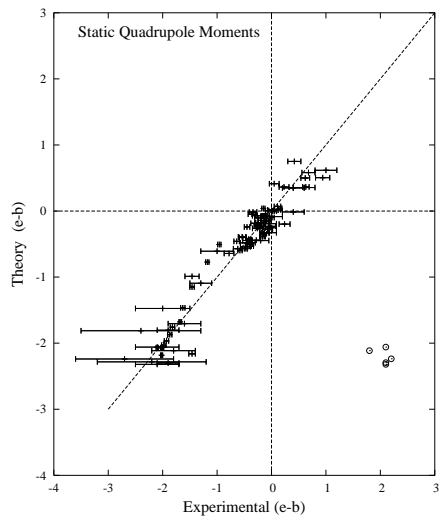}
\caption{Experiment compared to theory for the quadrupole moment of
\numstaticq
excited \twoplus states.  Experimental database is from the tabulation
in ref. \cite{st05}.
}
\label{fi:qstatic}
\end{figure}

Next we examine the spectroscopic quadrupole moments of the first excited
\twoplus states.  The comparison with experiment for \numstaticq nuclei
is shown in Fig.~\ref{fi:qstatic} on linear scales. For the
experimental data, we have used the tabulation by Stone in ref.
\cite{st05}. The quality of the experimental information is
quite variable, and we present only the cases where the assigned
error was much smaller than the magnitude of the moment.  Even
so, there are cases ($^{160}$Dy, $^{170,174,176}$Yb and
$^{180}$W) in which the sign of the
moment is not determined. These nuclei appear on the graph twice: once with error bars,
assuming that the theory gives the correct sign, and once with
open circles, assuming the opposite sign. They are all predicted
to be prolate with an negative quadrupole moment. Globally,
there is a very good agreement between $Q(2^+)$ measurements and
predictions.
   
\section{Performance by nuclear type}

It is clear that the 5DCH theory performs much better for
strongly deformed nuclei than for spherical ones.  To make this
observation quantitative, we attempted to sort the nuclei into
different categories and evaluate the $R_{Q,E}$ statistics by
category.  To make this as systematic as possible, we define the
categories on the basis of the theoretical properties of the
nuclei.  Thus, no empirical data is used to select the most
favorable cases.

The most obvious way to define a strongly deformed nucleus is to
make a cut on the mean quadrupole deformation $\bar{\beta}$.  However,
if the cut is at large enough $\bar{\beta}$ to include the well-known
deformed nuclei in the rare earths and actinides, it will also
keep certain light nuclei that have large mean $\bar{\beta}$ as well
as strong fluctuations in $\bar{\beta}$.  We therefore adopt the
classification by Sabbey, et al. \cite{sa06}, defining a
strongly deformed nucleus as one in which the mean deformation
$\bar{\beta}$ is larger than the r.m.s. fluctuation in $\bar{\beta}$.  We also
define a category ``semimagic" in which either the proton or
neutron number has the value $8,20,28,50,82,$ or 126.  The
remaining nuclei in our study can be considered either spherical
or soft-deformed; they are lumped together in the category
``other".  The $R_{Q,E}$ statistics by category are given in
Table \ref{ta}.  One sees that the bias in average energy
depends on the category: deformed nuclei are slightly
unpredicted, the ``other" category are overpredicted, and the
semimagic nuclei are correct on average.  However, the
dispersions in categories except the deformed one are large.
Here the semimagic nuclei are the poorest, with the theory too
high on average by almost 50 \%. In the bottom half of the
Table, for the reduced matrix elements, the entry for the
deformed nuclei stands out with an average of 0.035 and a
dispersion of less that 0.1.
\begin{table}[h!]
\caption{\label{energies}
Statistics for the performance of the 5DCH theory using the
Gogny D1S interaction on properties of \twoplus excitations. The quantities
$\bar{R}$ and $\sigma$ are defined in the text.
}
\label{ta}
\begin{tabular} {|cc|cc|}
\colrule
Category  & Number of Nuclei &   $\bar{R}$&  $\sigma$ \\
\colrule
\multicolumn{4}{|c|}{Excitation Energy}  \\
\colrule
All          &    519   &        0.12 &     0.33\\
Semimagic     &    73     &       0.02  &    0.51\\
Deformed&        146     &     -0.05   &   0.19 \\
Other    &       300     &          0.22&       0.29\\
\colrule
\multicolumn{4}{|c|}{\rme} \\
\colrule
All     &        319   &        0.10&     0.21 \\
Semimagic&        43     &      0.42 &    0.23\\
Deformed  &      106      &     0.035  &   0.09\\
Other       &    170         &   0.065     & 0.19\\
\colrule
\end{tabular}
\end{table}

\section{Conclusion}

In this work we have tested a theory of nuclear structure with
respect to the properties of the lowest \twoplus excitations
in even-even nuclei.  The theory is the microscopic collective
Hamiltonian for quadrupole degrees of freedom in which all parameters
are determined by CHFB input using the Gogny D1S interaction.  The
CHFB starting point imposes shell characteristics on the nuclear
structure, and this persists to a considerable extent in the
collective Hamiltonian.   The 5DCH theory is quite reliable for 
open proton and neutron shell nuclei.  Here there are strong
correlations that break the naive shell picture and often reorganize
the wave functions into a deformed band with distinctive signatures
in the properties of the first \twoplus excitation.  For nuclei
having only one open shell the theory is still useful for
predicting excitation energies and
quadrupole transition moments--much better than a factor 2--but one does not
achieve the quantitative,  10\%- level accuracy that we found for the
deformed nuclei.  Finally, in doubly closed shell nuclei, the
mapping of CHFB to the collective Hamiltonian breaks down and the
theory cannot be used.  Still, we found that the calculations
could be made applying the theory to more than 90 \% of the
nuclei whose \twoplus excitations are known.

For future work, the good results for deformed nuclei is encouraging
to a global study of other excitations implied by the 5DCH.
In particular, softness in the axial and triaxial coordinates $\beta$ and 
$\gamma$ gives rise to low-lying
vibrations.  The $\gamma$ vibrations 
appear as higher lying \twoplus excitations,
and the systematics of their energies and transition quadrupole
moments would provide a severe test of the 5DCH theory.  The
$\beta$ vibration appears as $0^+$ excitation on top of which
is built a rotational band; besides its energy in many cases its 
monopole transition matrix element to the ground state is known. 

Concerning the methodology, there are many points that
need more theoretical attention.  A weak point in the present
study is the 
Inglis-Belyaev approximation for the collective masses.
A better approximation would be that based on the QRPA theory~\cite{ri80}.
The same theory is also suitable for the prediction of collective and non-collective
modes in closed shell nuclei for which the GCM+GOA theory may break down.
Such works are in progress.

A systematic study of $2^{+}$ state energies has also been performed 
recently by Sabbey et al. ~\cite{sa06} using a different energy functional and methodology.
There i) the zero-range Skyrme force is used in Hartree-Fock+BCS
mean field calculations restricted to axial quadrupole deformation; 
ii) the configurations generated by the GCM are projected on
angular momentum J = 0 and 2 and good particle numbers; iii) the
configuration mixing is carried out directly rather than through a collective
Hamiltonian.
In principle that method is applicable to all nuclei including
the doubly magic, but they only reported results for two-thirds of the
measured nuclei, due to numerical difficulties associated with
the discrete basis. 

Their theory does very well on the quadrupole properties of deformed nuclei,
showing that this aspect of nuclear structure is robust in self-consistent
mean-field theory.   However, their calculated energies are systematically
higher than ours and higher than the data by about 50\%.  At this time
it is not clear what the origin of the difference is.  Certainly, the
dependence on the functionals and on the theoretical approximations needs
to be investigated further.

\section*{Acknowledgment}

This work was  supported  in part  by the Department of Energy Grants
DE-FG02-00ER41132 and DE-FC02-07ER41457.


\begin{thebibliography}{99}

\bibitem{ra02}
	S. Raman, C. W. Nestor, Jr., and P. Tikkanen, 
At. Data Nucl. Data Tables \textbf{78}, 1 (2001); the
\twoplus excitation energy of $^{114}$Ru has been corrected to the value
given in ref. \cite{sh94}.
\bibitem{ri80}
	P. Ring, P. Schuck, The Nuclear Many-Body Problem, Springer-Verlag, New York (1980) p. 424.
\bibitem{sa06} B.~Sabbey, M.~Bender, G.F. Bertsch, and P.-H.~Heenen, 
Phys. Rev. C \textbf{75} 044305 (2007).
\bibitem{sh94}  J.A.~Shannon, et al., Phys. Lett. B336 136 (1994).
\bibitem{li99}
	J.~Libert, M.~Girod, and J.-P.~Delaroche, Phys. Rev. C60 054301 (1999).
\bibitem{de06}
	JP Delaroche, et al., Nucl. Phys. A771 103 (2006).
\bibitem{go05}
	H.~Goutte,J.-F.~Berger, P.~Casoli, and D.~Gogny, Phys. Rev. C71 024316 (2005).
\bibitem{Bo88}
	W. Boeglin et al., Nucl. Phys. A477 (1988) 399.
\bibitem{Fl04}
  P. Fleischer, P. Kl{\"u}pfel, P.-G. Reinhard, and J. A. Maruhn,
  Phys. Rev. C \textbf{70} (2004) 054321.

\bibitem{de80}
	J. Decharg\'{e}, D. Gogny, Phys. Rev. C 21 (1980) 1568.
\bibitem{be91}
	J.-F. Berger, M. Girod, D. Gogny, Comp. Phys. Comm. 63 (1991) 365.
\bibitem{st05} 
	N.J. Stone, At. Data Nucl. Data Tables {\bf 90} 75 (2005).
\end{thebibliography}
\end{document}